\documentclass[twocolumn,preprintnumbers,amsmath,amssymb]{revtex4}

\usepackage{graphicx}
\usepackage{dcolumn}
\usepackage{bm}

\begin{document}

\title{Ge growth on ion-irradiated Si self-affine fractal surfaces }

\author{D. K. Goswami, K. Bhattacharjee and B. N. Dev}
\bigskip
 \altaffiliation{bhupen@iopb.res.in}
\affiliation {Institute of Physics, Sachivalaya Marg, Bhubaneswar -751005, India }

\date{\today}
\begin{abstract}

We have carried out scanning tunneling microscopy experiments under ultrahigh
vacuum condition to study the morphology of ultrathin Ge films deposited on
pristine Si(100) and ion-irradiated Si(100) self-affine fractal surfaces. The
pristine and the ion-irradiated  Si(100) surface have roughness exponents of
$\alpha=0.19\pm0.05$ and $\alpha=0.82\pm0.04$ respectively. These measurements
were carried out on two halves of the same sample where only one half was 
ion-irradiated. Following deposition of a
thin film of Ge ($\sim6\AA$) the roughness exponents change to 0.11$\pm$0.04
and 0.99$\pm$0.06, respectively. Upon Ge deposition, while the roughness increases by more than an
order of magnitude  on the pristine surface, a smoothing is
observed for the ion-irradiated surface. For the ion-irradiated surface the
correlation length $\xi$ increases from 32 nm to 137 nm upon Ge deposition. Ge
grows on Si surfaces in the Stranski-Krastanov or layer-plus-island mode where
islands grow on a wetting layer of about three atomic layers. On the pristine
surface the islands are predominantly of square or rectangular shape, while on
the ion-irradiated surface the islands are nearly diamond shaped. Changes of
adsorption behaviour of deposited atoms depending on the roughness exponent (or
the fractal dimension) of the substrate surface are discussed.\\


\end{abstract}
 \maketitle
\section{Introduction}

A wide variety of surfaces are present in nature where surface roughness is
well  described in terms of self-affine fractal scaling spanning various
length scales $-$ for example, the kilometer-scale structure of mountain terrain
\cite{mandel} to nanometer-scale topology of thin films obtained by deposition of atoms
or molecules on substrates \cite{chiarello} and of ion-bombarded surfaces 
\cite{eklund,krim,dipak1}. Various physical processes including
fracture, ion bombardment, molecular beam epitaxy etc. produce this kind of
surface morphology. In thin film deposition the morphology of the bare surfaces 
of substrates prior to thin film
deposition can also influence the morphology of the deposited films. To a large
extent the usefulness of thin films in terms of mechanical, optical and 
electrical properties depends on the surface
morphology. It is important to understand growth mechanisms leading to various
surface morphologies.

Ion irradiation and molecular beam epitaxy (MBE) $-$ both used in 
advanced modern technology $-$ affect surface morphologies in different ways.  In the
case of heteroepitaxial MBE growth there are three different growth modes 
$-$ layer-by-layer (Frank-van der Merwe or FM), island (Volmer-Weber or VW) and layer-plus-island  
(Stranski-Krastanov or SK) \cite{bauer}. Thus MBE growth
can produce different surface morphologies.  Which growth mode will be adopted
in a given system depends on the surface free energy of the
substrate, free energy of the film, their interface energy and the strain 
energy due to lattice mismatch. Thus the surface morphology is
governed by these parameters. Ion bombardment produces rough surface
as well as smooth surface under different experimental conditions. A common
feature of most of the ion-bombarded surfaces is self-affine fractal structure.
The fractal dimensions of such self-affine surfaces can be controlled. The
adhesion of a thin film deposited on a fractal surface is expected to be
affected by the fractal dimension of the substrate surface \cite{pla}. This inspires
studies of the morphology of MBE-grown thin films on ion-irradiated substrates. Here
we report our results of such studies and compare the growth on pristine and 
ion-irradiated surfaces. 
 
Ion-solid interaction alters the topography of solid surfaces via competing
surface roughening and smoothing processes. These competing processes are
responsible for the creation of surface features like quasiperiodic ripple
\cite{chason,carter,wittmaack,habenicht} and self-affine fractal topographies
\cite{habenicht,eklund,krim,dipak1}. In the ion mass-energy regime where sputtering is
dominant, surface roughening is observed \cite{habenicht,eklund,krim}. 
When ion-beam induced effective surface
diffusion dominates surface smoothing is observed \cite{dipak1,carter,bob}.  
The smoothed surface can also be a self-affine fractal surface \cite{dipak1}.

Surface morphology  is often described by some statistical
parameters. Most common are the surface width ($\sigma$), represented by the
root-mean-square roughness value, and the in-plane correlation length ($\xi$). 
The correlation length is the average distance between features in the surface profiles within which
the surface variations are correlated. Surface width is an important parameter
that represents a measurement of the correlations along the direction of the
surface growth. Scaling studies can be performed by measuring
surface roughness at various length scales. Root-mean-square surface roughness
$\sigma$ is defined as 
\begin{equation}
\sigma=<[h(x,y)-h]^2>^{1/2},
\end{equation} 
where $h(x,y)$ is the
surface height at a point $(x,y)$ on the surface and $h$ is the average height.
If the horizontal sampling length on the surface is $L$ and $\sigma\propto
L^\alpha$, where $0<\alpha<1$, the surface is termed self-affine \cite{krim}.
The value of the scaling exponent $\alpha$ indicates how roughness changes with
length scales. However, it does not tell whether the surface is roughened or
smoothed upon any kind of surface treatment. Small $\alpha$ values are
associated with jagged surfaces (anticorrelation) while large values indicate
well correlated, smoothed-textured surfaces.  The surface width (rms roughness)
scales with linear length $L$ for $L<<\xi$ and for $L>>\xi$ rms roughness
$\sigma(L) \rightarrow \sigma_0$. For the case $\xi>>a$ (lattice spacing of the
system) the analytical form of roughness can be written as \cite{palasantzas}:
\begin{equation}
\sigma^2(L) = \frac{\sigma_0^2}{(1+4\pi\xi^2/2\alpha L^2)^\alpha}
\end{equation}
In ion-surface interaction, in a previous work we observed ion-beam induced
surface smoothing and the smoothed surface to be self-affine.
2.0 MeV Si$^+$ ions bombarded on a Si(100) surface (hereafter denoted IB) with the fluences in the range
$10^{15}-10^{16}$ ions/cm$^2$, produced a self-affine fractal surface with a scaling exponent
$\alpha = 0.53\pm0.03$ at length scales below 50 nm \cite{dipak1}.
 In ion-atom collisions in solids and at
the surface, the elastic energy lost by an ion is transferred to a recoil atom,
which itself
collides with other atoms in the solid and so forth. In this way the ion
creates what is called a collision cascade. The displaced atoms in this
collision cascade may acquire a kinetic energy enough  to escape from the solid
surface $-$ a phenomenon known as sputtering. However, if the energy (component
normal to the surface) of the displaced atoms is smaller than the surface
binding energy, the atoms may reach the surface but cannot leave the surface.
They can, however, drift parallel to the surface providing an effective
surface diffusion. Apparently this ion-beam induced effective surface diffusion
causes nanometer-scale surface smoothing \cite{dipak1}. Discovery of this
ion-irradiation induced nanoscale surface smoothing phenomenon leading to a
self-affine fractal surface inspired further investigations \cite{dipak2}
including the present one.

In surface studies and epitaxial thin film growth, thermal treatment of the solid substrate
is almost inevitable. Thermal treatment of ion-bombarded surfaces  (hereafter denoted TIB)
produced a self-affine surface with a different roughness exponent with both
smoothing and roughening below and above a particular length scale,
respectively. At length scales below $\sim50$ nm, thermal treatment  was found to cause
further smoothing of the ion-bombardment induced smooth
surface (IB).  However, at length scales above $\sim50$ nm the surface roughness
 of the TIB samples increases for lateral dimensions  upto $\sim300$ nm with a scaling
exponent $\alpha = 0.81\pm$0.04 over the entire range covering both smoothing and roughening regimes 
\cite{dipak2}. Scaling studies for Ge growth
on this surface as well as on the pristine Si(100) surface  are reported here.

Another motivation for the study of growth of Ge on ion-irradiated Si surfaces
is related to our earlier studies of growth of Ge nanostructures on silicon
\cite{amal1,amal2}. Growth of Ge nanodots and nanowires on polymer-coated Si
surfaces indicated that nanodots are arranged on defects, which might be
utilized to fabricate lattices of nanodots. Patterned defect structures can be
created on Si by ion-irradiation. Ge growth on this patterned surface could
possibly be used to form Ge nanodot lattices. However, these aspects will not
be presented here.

\section{Experimental}

Si(100) substrates with the native oxide were irradiated with 2.0 MeV
Si$^+$ ions. The ion beam was incident along the surface normal ($\theta\approx0^\circ$)
 and rastered on the sample in order to obtain a uniformly
irradiated area. One half of the sample was masked and hence unirradiated. An
ion beam flux of $\approx1\times10^{12}$ cm$^{-2}$ sec$^{-1}$ was used with a
fluence of $4\times10^{15}$ ions/cm$^2$. The pressure in the
irradiation chamber was $\sim 10^{-7}$ mbar. Following irradiation the sample
was taken out and inserted into an ultrahigh vacuum (UHV) chamber (pressure:
3$\times$ 10$^{-10}$ mbar) containing an Omicron variable temperature scanning
tunneling microscope. Scanning tunneling microscopy (STM) measurements were performed at room
temperature.  The roughness measurements were made on both the pristine as
well as the irradiated half of the sample with a thin ($\sim$ 1.5 nm) native oxide
layer.  For preparing a clean silicon surface, the sample was degassed about
600$^\circ$C for 12 hours prior to the flashing at 1200$^\circ$C for 2$-$3 minutes under UHV
condition (1$\times 10^{-10}$ mbar) in a molecular beam epitaxy (MBE) growth chamber.
This process removes the native oxide and exposes a clean Si surface. The MBE and the STM chambers are connected. This system is described in ref \cite{dipak3}
On the clean pristine half of the sample we
observe surface atomic steps as usually observed on the
atomically clean Si(100) surfaces. Roughness measurements were again made on
both the pristine and the ion-irradiated halves of the sample after removal of the oxide. 
Roughness exponents were determined from STM images. A large number of scans, each of size $L$, were
recorded on the surface at random locations. The $\sigma$ values for the rms
roughness given by the instrument plane fitting and subtraction procedure had
been carried out. This procedure was repeated for many different sizes and a
set of average $\sigma$ versus $L$ values was obtained. On the entire clean (oxide removed)
surface, having performed scaling measurements, $\sim6 \AA$ Ge was
deposited while the substrate was kept at $550^\circ$C (the standard condition for
Ge epitaxy on silicon). Scaling measurements on both halves have been made again
following Ge deposition. 
\begin{figure}
\includegraphics[height=7cm]{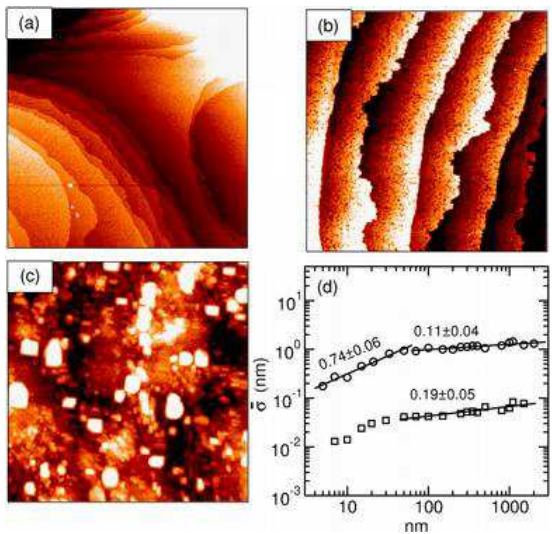}
\caption{ STM images (a) ($1500\times1500$ nm$^2$) and (b)
($300\times300$ nm$^2$) of a clean pristine Si(100) surface. Atomic steps and
terraces are seen. (c) A STM image (400$\times$400 nm$^2$) of a Ge-deposited
film on Si(100). (d) Average rms surface roughness ($\bar\sigma$) vs scan size
($L$) on the pristine Si(100) [$\square$,TP] and on the Ge-deposited pristine
surface [{\LARGE$\circ$},TP+Ge] surfaces. The values of the roughness exponents
$\alpha$ are shown in Fig.1(d). }
\end{figure}

\section{Results and discussions}

\subsection{Ge deposition on the pristine surface}

The pristine half of the Si(100) surface shows the typical (2$\times$1) reconstruction
[This sample will be referred to thermally treated pristine, TP]. 
STM images [Fig.1(a),(b)] show flat terraces and atomic steps. Ge deposition on
the pristine surface leads to Ge island growth [Fig.1(c)]. Ge is known to grow
on Si in the Stranski-Krastanov or layer-plus-island mode.  The results of
roughness scaling studies on the clean Si(100) surface as well as on the 
Ge-deposited surface are shown in Fig.1(d). For length scales $>$50 nm roughness
exponents are found to be $0.19\pm0.05$ and $0.11\pm0.04$ for the clean Si(100) and the
Ge-deposited Si(100) surface, respectively. $\alpha\rightarrow0$ corresponding to 
local dimension D=3$-\alpha\rightarrow3$, is a subtle situation, since a three
dimensional object can be either a fractal or a volume. As we notice, clean
surface is dominated by (100) terraces with monatomic steps [Fig.1(b)]. At
larger length scales multiple steps with short terraces or step bunching are
encountered; this tends to increase the roughness values. Roughness values are
very small ($\sim0.05$ nm). Upon Ge deposition, the value of roughness
increases more than an order of magnitude, however the roughness exponent does
not change significantly. Roughness exponent observed here is quite small.
Observation of such a small value of $\alpha$ ($0.12\pm0.05$) was earlier
reported for roughness on a Ag film deposited by thermal evaporation on a
polished quartz crystal \cite{palasantzas}. For the Ge film, surface is smoother at length
scales $<$50 nm with the roughness exponent $\alpha=0.74\pm0.06$. Assuming that the knee
regime in the $\sigma(L)$ plot corresponds to a length scale approximately
4$\xi$, for the Ge-deposited surface the in-plane correlation length $\xi$ is
$\sim$12 nm.
\begin{figure}
\includegraphics[height=7cm]{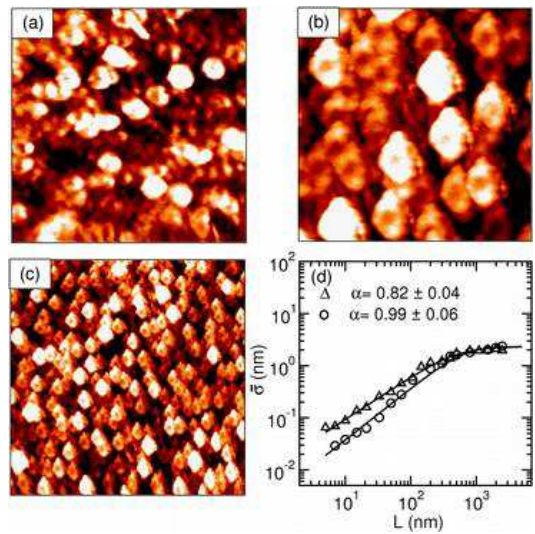}
\caption{
(a) A STM image ($1100\times1100$ nm$^2$) of an ion-irradiated
cleaned (TIB) Si(100) surface. STM images (b) ($1100\times1100$ nm$^2$) and (c)
($3000\times3000$ nm$^2$) from the Ge-deposited (TIB+Ge) surface. (d) Average
roughness ($\bar\sigma$) vs scan size ($L$) plots: for the TIB [$\triangle$]
surface and the TIB+Ge [{\LARGE$\circ$}] surface. Roughness exponents are
marked in the figure.
}
\end{figure}
\subsection{Ge deposition on the ion-irradiated surface}

Surface morphology of an ion-irradiated clean Si(100) surface is seen in the STM image of
Fig.2(a). Fig.2(b) and Fig.2(c) show STM micrographs following Ge deposition.
These are quite different from that for Ge deposition on a pristine Si(100)
surface. Nearly diamond shaped islands are seen in Fig.2(b) and Fig.2(c), while
dominant square or rectangular shaped islands are seen in Fig.1(c). An ideal
Si(100) surface has a fourfold symmetry, while the ($2\times1)$ reconstructed
Si(100) surface has a twofold symmetry. These dictate the square or rectangular
island shape. On the irradiated clean Si(100) surface, neither large terraces
nor (2$\times$1) reconstruction is observed.
The log-log plots of average surface roughness ($\bar\sigma$) versus scan size
$L$ are shown in Fig.2(d).  Above the length scales
of $\sim$700 nm both the surfaces practically have the same roughness. That is, the Ge
deposition has hardly any effect  on the saturation roughness, $\sigma_0$,
at these length scales. However, for length
scales below 700 nm, Ge-deposited surface shows considerable surface smoothing
than the TIB surface. From the linear portion of the log $\bar\sigma$ versus
log $L$ plot the roughness scaling exponents have been found as 0.82$\pm$0.04
for the TIB surface and 0.99$\pm$0.06 for the Ge-deposited surface. This
indicates that both the surfaces are self-affine fractal in nature. 
\begin{table}
{\begin{minipage}{3.0in}
\caption {Roughness exponent ($\alpha$) and the in-plane correlation length ($\xi$)}
\end{minipage}}
\vspace*{0.5cm}
\begin{tabular}{ p{1.0in}| p{0.8in}| p{0.8in} } \hline\hline
Sample & ~~~~~~$\alpha$ & ~~~~~~$\xi$ (nm)\\ \hline
TP Si(100) &~~0.19$\pm$0.05 & ~~~~~~$-$\\
TP Si(100)+Ge  &~~0.11$\pm$0.04 & ~~~~~~$-$\\
{$ $} &~~0.74$\pm$0.06 & ~~~~$\sim12$\\
IB Si(100) & ~~0.53$\pm$0.03\footnote{from Ref.\cite{dipak1}} & ~~~~~~$-$\\ 
TIB Si(100) & ~~0.82$\pm$0.04\footnote{from Ref.\cite{dipak2} and the present study} & ~~~~32$\pm$3.0 \\
TIB + Ge  & ~~0.99$\pm$0.06 & ~~~~137$\pm$10.0 \\ \hline\hline
\end{tabular}
\end{table}

Fits using Eq.(2) are also shown in Fig.2(d). We have used the values of
$\alpha$ in Eq.(2) from the fitting of the linear portion of the curve in the
log-log plot. The results are presented in  Table-1.

As discussed in ref\cite{dipak2} thermal treatment of the ion-irradiated Si(100) surface (TIB)
simultaneously smoothes and roughens the surface at different length scales,
with the same roughness exponent over the entire length scales. Ge deposition
causes further smoothing over this entire region [Fig.2(d)] with an increased in-plane
correlation length. The correlation length $\xi$, together with the roughness
exponent $\alpha$, controls how far a point on the surface can move before losing
the memory of the initial value  of its
height ($h$) coordinate. 

After removal of the native oxide by thermal treatment, the roughness of the
irradiated (TIB) surface is found to be much larger compared to the pristine
(TP) surface. The reason for this has been explained in ref.\cite{dipak2} by Monte Carlo
simulation results showing atomic displacements at the ion-bombarded oxide/Si
interface. The loss of symmetry and the new structure formation on the TIB
surface is apparently responsible for the difference in morphology of the Ge-deposited
surfaces in Fig.1(c) and Fig.2(b) [also Fig.2(c)]. The difference in
morphologies of the clean surfaces as seen in Fig.1(a) [also Fig.1(b)] and Fig.2(a) would lead
to a difference in adhesion of deposited atoms on them. The smoothing observed
at smaller length scales upon Ge-deposition on the TIB surface indicates that
the Ge atoms fill the surface troughs, where the deposited atoms would have
more nearest neighbors for better bonding. [For length scales $\le 200$ nm the
Ge-deposited TIB surface is also smoother than the Ge-deposited TP surface].
The theoretical work of Palasantzas and Hosson \cite{pla} predicting that the adhesion of a
thin film deposited on a fractal surface would depend on the fractal dimension
of the substrate surface needs to be further explored through experimental
investigations.

Considering the fact that the ion-irradiation was performed at $\sim10^{-7}$ mbar, 
one may wonder about deposition of C on the irradiated surface and its
influence on subsequent results. However, we do not expect any significant C
deposition. This aspect has been explained in details in ref. \cite{dipak1}.

\section{Summary and conclusions}

Roughness scaling behavior of thin Ge films on ion-irradiated and pristine
Si(100) surfaces has been investigated. Ge was deposited after removal of the
native oxide under ultrahigh vacuum condition. On the pristine surface, upon Ge
deposition, although the roughness increases more than an order of magnitude the
roughness exponent $\alpha$ changes from 0.19 to 0.11. On the ion-irradiated surface, Ge
deposition causes surface smoothing with roughness exponent changing from 0.82
to 0.99. In these two cases, different fractal dimensions of the fractal
surfaces of the substrates, 2.81 for the pristine and 2.18 for the ion-irradiated, has apparently led to
different surface morphologies of the Ge-deposited surfaces. Ge islands on the
pristine surface tend to have square or rectangular shape, while the islands on
the ion-irradiated surface have nearly diamond shaped structures. For metal
films on semiconductors, recently interesting island structures with quantized
heights $-$ apparently an effect of electronic confinement $-$ have been observed
\cite{gavioli,dipak4,su}. This type of growth cannot be explained within the
traditional FM, SK and VW growth modes. Recent theoretical models indicate
an {\it electronic growth} mechanism for such systems, where energy
contribution of the quantized electrons confined in the metal overlayer can
actually determine the morphology of the growing film \cite{okomoto}. The
electronic structure of these islands would depend on their shape. The
possibility of influencing the shape of the islands by growth on surfaces of
different fractal dimensions would provide the capability of manipulating the
electronic behavior of such nanostructures.

\newpage

 \end{document}